\documentclass[twocolumn,amsmath,amssymb,floatfix]{revtex4}

\usepackage{bm}
\usepackage{amsmath}
\usepackage{epsf}

\def\simlt{\lesssim}

\newcommand{\ApJ}{Astrophys. J}

\newcommand{\PRD}{Phys. Rev. D}
\newcommand{\MNRAS}{Mon. Not. R. Astron. Soc.}
\newcommand{\ARAA}{Ann. Rev. Astron. Astrophys.}

\newcommand{\aut}[2]{{#2.\ #1,}}
\newcommand{\saut}[2]{{#2.\ #1,}}
\newcommand{\laut}[2]{{#2.\ #1,}}
\newcommand{\refs}[6]{#2, {\bf #3},  {#4} (#5).}
\newcommand{\mrefs}[6]{#2, {\bf #3},  {#4} (#5);}

\newcommand{\mybib}[2]{\bibitem{#2}}

\begin{document}

\title{Model-Independent Reionization Observables in the CMB}
\author{Wayne Hu$^{1}$ and Gilbert P. Holder$^{2}$}
\affiliation{
{}$^{1}$Center for Cosmological Physics, 
Department of Astronomy and Astrophysics, 
and Enrico Fermi Institute, University of Chicago\\
{}$^{2}$School of Natural Sciences, Institute for Advanced Study
}

\begin{abstract}
\baselineskip 11pt We represent the reionization history of the
universe as a free function in redshift and study the potential 
for its extraction from CMB polarization spectra.
From a principal component analysis, we show that the ionization 
history information is contained in 5 modes, 
resembling low-order Fourier modes in redshift space.
The amplitude of these modes represent a compact
description of the observable properties of reionization in the CMB,
easily predicted given a model for the ionization fraction.
Measurement of these modes can ultimately constrain the total 
optical depth, or equivalently the initial amplitude
of fluctuations to the 1\% level regardless of the true model
for reionization.  
\end{abstract}
\maketitle

\section{Introduction}
\label{sec:intro}

The WMAP experiment \cite{WMAP} has recently
detected the reionization of the universe through the large-angle
polarization of the cosmic microwave background (CMB).
Rescattering of CMB radiation bearing a quadrupole temperature
anisotropy leads to a small linear polarization which is then
correlated with the temperature anisotropy itself.
It was through this cross-correlation that WMAP measured the
total optical depth to Thomson scattering, $\tau =0.17 \pm 0.04$ \cite{WMAP}.

The CMB power spectra potentially contain more information than
the optical depth integrated over the whole ionization history \cite{Kap03}.
Unfortunately, the nature of the ionizing sources is not currently 
well-understood
(e.g.\ \cite{BarLoeb01}, c.f.\ \cite{Gne00}), and therefore the detailed ionization history should
not be treated in the traditional CMB approach of adding well-motivated
model parameters to a parameter estimation chain.  
Attempts to do so can result in biases and 
conflicting results.
For example, using an ionization history derived from a particular
numerical prescription \cite{BruFerScan02} it was found that 
the partially ionized epoch was undetectable in the CMB, while
analyses using semi-analytic models \cite{WyiLoeb03,models}
showed that the predicted multiple
epochs of reionization and partial recombination can be detected 
\cite{NasChi03,HolHaiKapKno03}.
As another example, by assuming an 
overly simplistic step function model for reionization, the total optical depth
$\tau$ in the semi-analytic models can be mis-estimated by up to a 
few $10^{-2}$ \cite{HolHaiKapKno03};
while not a large amount it is already approaching the statistical errors
in the first year WMAP data.

These uncertainties argue for a more model-independent approach 
to the phenomenology of reionization in the CMB.  In this paper,
we begin by considering a complete basis for the ionization history
in \S \ref{sec:model}.  In \S \ref{sec:info}, we employ a principal
component analysis to isolate the CMB reionization observables. 
Throughout we work in 
a flat $\Lambda$CDM cosmology with 
$h=0.72$, $\Omega_m h^2=0.145$, $\Omega_b h^2=0.024$ and $n=1$. 
In the absence of systematic errors, most of the ionization history information 
comes from the power spectrum of the $E$-mode polarization $C_\ell^{EE}$ 
as opposed to the temperature-polarization cross correlation measured by WMAP
\cite{TegEisHudeO00} and for simplicity we will here consider
only this power spectrum. Nevertheless the main results are applicable to the
cross power spectrum as well.

\section{Model-Independence}
\label{sec:model}

To zeroth order, the total Thomson optical depth parameterizes
the effects of reionization independently of the specific model
for the ionization history. 
Firstly, the acoustic peaks in 
the CMB power spectra are lowered by $e^{-\tau}$ in amplitude.
Because the effect 
is degenerate with a change in the initial normalization of the curvature
fluctuations $\delta_{\zeta}$,
it is absorbed into fixing the combination $\delta_{\zeta} e^{-\tau}$.   
Secondly, rescattering generates  large angle temperature
fluctuations through the Doppler effect and thirdly, it
generates large angle polarization.  Both effects have an  rms 
that depends mainly on $\tau$ but since the primordial polarization
is free of large angle contributions, it is the better probe of
reionization. 
Finally there are further signals from inhomogeneous scattering
that appear beyond the damping tail ($\ell\gtrsim 2000$) but 
whose phenomenology is
inextricably tied to the model for the ionizing sources;
we will not consider them further here.

\begin{figure}[tb]
\centerline{\epsfxsize=3.25truein\epsffile{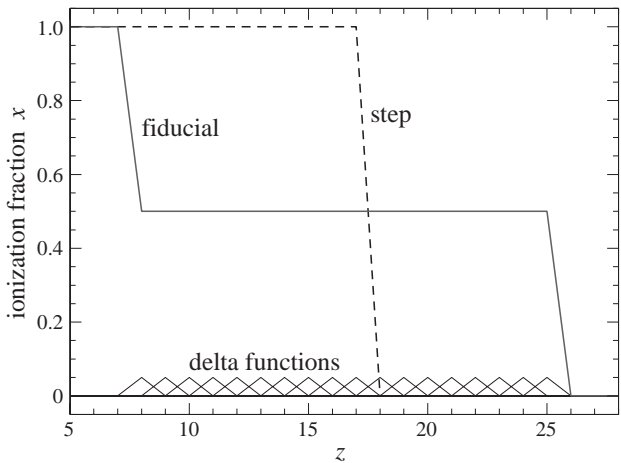}}
\caption{Hydrogen ionization fraction $x$ as a function of redshift $z$
in the fiducial model (thick), traditional step function 
ionization (dashed) and delta-function perturbations (thin).}
\label{fig:xe}
\end{figure}

In detail, the residual temperature and polarization effects
depend on the reionization history, mainly through a sensitivity to 
the horizon scale at the epoch of rescattering \cite{coherence}.  
We assume the hydrogen 
ionization fraction $x$ to be a free function of redshift.  A complete basis
for this function can be formed from a series of delta functions 
in redshift.  We approximate this series as Kronecker delta
functions $\delta x(z_i)$ spaced by $\Delta z$ with linear interpolation
between the points (see Fig.~\ref{fig:xe}).  The spacing
is chosen to capture all of the information in the CMB and
also be sufficiently fine to reproduce features in the ionization model.
We take $\Delta z=1$ throughout. 

\begin{figure}[tb]
\centerline{\epsfxsize=3.25truein\epsffile{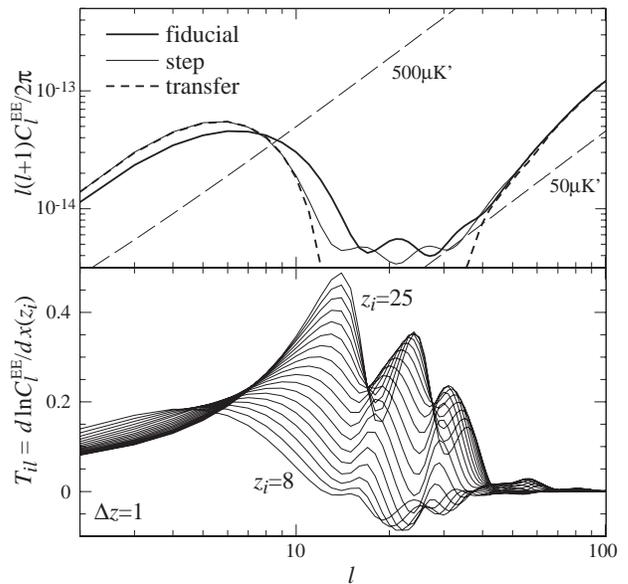}}
\caption{Top: $E$-mode polarization power spectrum for: 
the fiducial model of Fig.~\ref{fig:xe} (thick);  the step function
model (thin); the step function model with deviations transferred
onto the fiducial model (dashed); instrumental noise $w_P^{-1/2}$ 
(denoted in $\mu$K-arcmin) 
that roughly brackets expectations from WMAP and Planck 
(long dashed).
Bottom: the transfer function 
or fractional power spectrum response to a delta function 
perturbation of unit amplitude at $8\le z_i \le 25$.}
\label{fig:cl}
\end{figure}

A specific model can then be represented as a sum over the
delta function modes in the approximation of linear interpolation
between the points.  
Consider a fiducial ionization history with a hydrogen ionization
fraction of  $x=1$, $z_i \le 7$; $x=0.5$, $7 < z_i < 26$; $x=x_{\rm rec}$,
$z_i \ge 26$ 
where $x_{\rm rec} (\approx 0)$ is the ionization fraction coming out
of standard recombination  (see Fig.~\ref{fig:xe}).
Compare that to a step function 
ionization with $x = 1$, $z_i \le 17$ and $x_{\rm rec}$ $z_i > 17$ 
(Fig.~\ref{fig:xe}).  Both models have $\tau \approx 0.17$ 
but differ in the coherence scale of the polarization.  
These two models are clearly distinguishable in principle 
(see Fig.~\ref{fig:cl}, top)
and so more information than the optical depth can be extracted 
from precise measurements \cite{Kap03,HolHaiKapKno03}.

The parameters $x(z_i)$ may be simply appended to the usual
CMB parameter estimation chain but that approach has 
several drawbacks.  The $\Delta z$ spacing of the points
is arbitrary and moreover for $\Delta z=1$
the parameters are highly degenerate and would cause numerical
problems in a likelihood analysis.  To see this, consider
the effect of a delta-function perturbation to the fiducial
ionization history given above and quantify the response
in the power spectrum as a transfer matrix 
\begin{equation}
T_{\ell i} \equiv {\partial \ln C_\ell^{EE} \over \partial x(z_i)}\,,
\end{equation}
where we vary $7 < z_i < 26$ as motivated by the 
range in redshift over which reionization is expected 
to occur \cite{BarLoeb01}.   
The derivative is calculated by a double-sided finite difference
of $\delta x=0.05$.  
This matrix may be viewed as a transfer function for 
small perturbations from
the fiducial model.   The main feature is that perturbations at
relatively low redshift appear at low $\ell$ and those at high
redshift at high $\ell$.  The perturbations also introduce
ringing into the power spectrum which is retained in the
full model.  Because neighboring redshift modes produce similar
responses in the power spectrum, there exists a degeneracy
in the recovery of $\delta x(z_i)$.  

Under the linear approximation, we can pose the inverse problem 
as the recovery of $\delta x(z_i)$ from the linear response
\begin{equation}
\delta C_\ell^{EE} 
= C_\ell^{EE} \Big|_{\rm fid}  \sum_i T_{\ell i} \delta x(z_i)
+ N_\ell\,,
\end{equation}
where
$N_\ell$ represents noise sources.  
Instrumental noise levels that roughly bracket 
the expectations for WMAP and
Planck \cite{planck} are shown in Fig.~\ref{fig:cl} but systematic
effects such as residual foreground contamination will likely
dominate the errors.
With a characterization of the
statistical properties of the noise and a regularization of the reconstructed
signal, e.g. by placing a smoothness criteria on $x(z_i)$ through a two-point 
prior, standard techniques such as Wiener filtering
allow for model reconstruction.  It is important to take a fiducial
model that is close to the observed power spectrum since linear response
will only hold for small perturbations.  It will also be important
below in that the cosmic variance of the
assumed signal sets the fundamental noise in the
reconstruction.  We chose a fiducial model with an ionization 
$x(z_i)=0.5$ in the parameterized regime so that, 
neglecting helium reionization \cite{WyiLoeb03}, the maximal
excursion from the model is $\delta x(z_i)=0.5$. 

As a worst case scenario consider the reconstruction of the step 
function ionization in Fig.~\ref{fig:xe} from the fiducial model,
$\delta x(z_i)$ and the transfer matrix.  
Here all of the parameters $x(z_i)$ are offset 
from the fiducial choice by the maximum value.   The predicted power spectrum 
is shown in dashed
lines in Fig.~\ref{fig:cl} (top).  Note the agreement is still quite
good at the peak of the power but fails in the low signal regime
where small errors in the transfer function lead to large fractional
effects.  Still, given realistic measurement noise and 
an iterated choice of fiducial
model, the linear response approximation 
can provide useful tools for representing the
data.  A detailed consideration is beyond the scope of this work.  Instead
we will show that a modification of the forward approach 
suffices to extract essentially all of the information in the CMB.

\begin{figure}[tb]
\centerline{\epsfxsize=3.25truein\epsffile{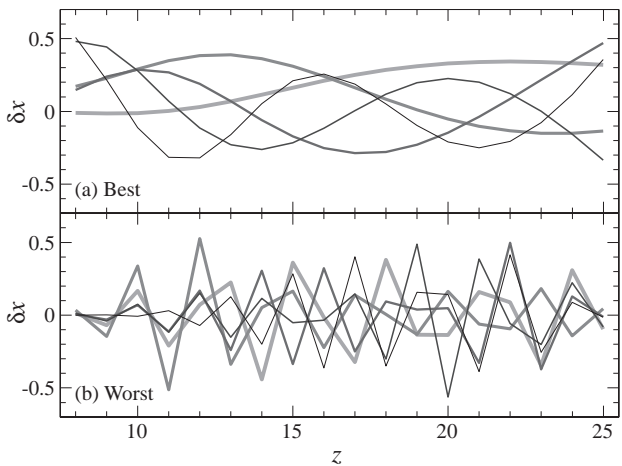}}
\caption{Eigenmodes. (a) 5 best (decreasing eigenvalue thick to thin) 
constrained eigenmodes or linear combinations of ionization history.
(b) 5 worst constrained eigenmodes.}
\label{fig:eigen}
\end{figure}

\section{CMB Observables}
\label{sec:info}

To better quantify the information contained in the power spectrum,
let us consider the ultimate limit of an all-sky experiment that
is cosmic variance limited.   The variance of the power spectrum
is then given by
\begin{equation}
\left< \delta C_\ell^{EE} \delta C_{\ell'}^{EE} \right> 
= {2 \over 2\ell+1} (C_l^{EE})^2 \delta_{\ell \ell'}
\end{equation}
and hence the covariance of the ionization parameters
$\left< \delta x(z_i) \delta x(z_j) \right> \approx ({\bf F}^{-1})_{ij}$,
where
\begin{equation}
F_{ij} = \sum_\ell (\ell+1/2) T_{\ell i} T_{\ell j}
\end{equation}
is the Fisher matrix.  The structure of the transfer matrix 
implies a large covariance between 
estimates of $\delta x(z_i)$ and renders the delta-function representation
difficult to visualize.

\begin{figure}[tb]
\centerline{\epsfxsize=3.00truein\epsffile{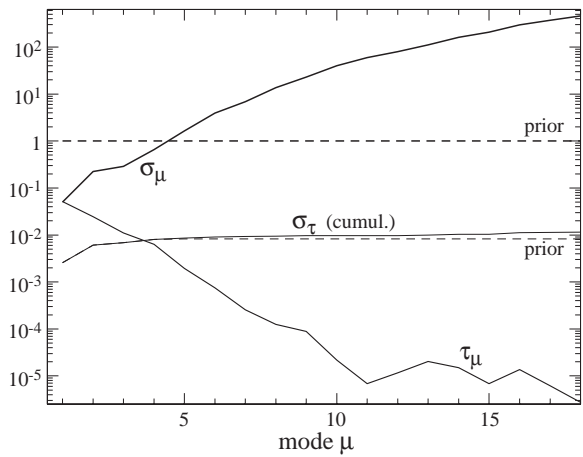}}
\caption{Eigenmode statistics. Top curve: rms error $\sigma_\mu$
on mode amplitude; dashed line represents a physicality prior on $x$; only the first 5 modes 
contain interesting information.  
Bottom curve: optical depth per unit-amplitude mode $\tau_\mu$. 
Middle curve: rms error on total optical depth shown as the cumulative
contribution from modes $\le \mu$; dashed line represents the physicality prior
on $x$.}
\label{fig:mode}
\end{figure}

Consider instead the principal component representation based
on the orthonormal eigenvectors of 
the Fisher matrix, decomposed as 
\begin{equation}
F_{ij} = \sum_{\mu} S_{i\mu} \sigma_{\mu}^{-2} S_{j \mu}\,.
\end{equation}
For a fixed $\mu$, the $S_{i\mu}$ specify linear combinations of 
the $\delta x(z_i)$ for a new representation of the data
\begin{equation}
m_\mu  =  \sum_i S_{i\mu} \delta x_i\,, 
\end{equation}
where the covariance matrix of the mode amplitudes
is given by
\begin{equation}
\left< m_\mu m_\nu \right> = \sigma_\mu^2 \delta_{\mu \nu}\,.
\end{equation}
In other words,
the eigenvectors form a new basis that is complete and yields uncorrelated
measurements with variance
given by the inverse eigenvalue.
The largest eigenvalues
correspond to the minimum variance directions and the first 5 are shown in
the upper panel of Fig.~\ref{fig:eigen}.  The first two 
correspond essentially to the average ionization at high
redshift and low redshift respectively.  
The lower panel shows the directions with the 5 highest variances.  Here
neighboring delta modes with similar responses compensate each other to leave
the observable power spectrum unchanged. The rms of each mode is shown
in Fig.~\ref{fig:mode}.  Because the ionization fraction cannot be
negative and the amplitude of each mode is $\sim 0.5$, only the first
5 modes with $\sigma_\mu \simlt 1$ have useful information. 
An added benefit of the principal component representation is that
the structure in the lowest modes is invariant under refinement of 
the binning scheme $\Delta z$.  In $\ell$ space, the first mode controls
the high $\ell$ power, the second the low $\ell$ power and the third
through fifth adjust the ringing in the spectrum.

\begin{figure}[tb]
\centerline{\epsfxsize=3.25truein\epsffile{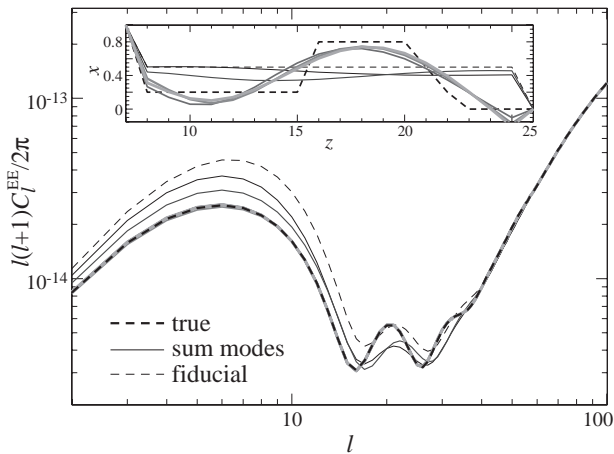}}
\caption{Representation of an arbitrary ionization history with the
first 5 eigenmodes.  Inset: ionization history in the true model
(thick dashed) compared with representation with 1 to 5 eigenmodes (solid,
increasing thickness) away from the fiducial model (thin dashed).  
Main panel: resulting prediction for the power spectrum.  With three or 
more modes the prediction is indistinguishable from the true model.}
\label{fig:clmode}
\end{figure}

These eigenmodes provide a good meeting ground between
observations and models.  
The amplitude $m_\mu$ of these 5 best modes may be added to the usual
CMB parameter estimation chain and the results compared to model predictions
for $m_\mu$ without significant loss of information.  As an example,
in Figure \ref{fig:clmode} we have represented a complex ionization
history (inset, thick-dashed) through 
its first 1 through 5 eigenmodes.  For this
ionization history the first 3 eigenmodes suffice to recover the observable
power spectrum.  
The temperature polarization cross spectra converges similarly.
Note that in the eigenmode analysis the assumed ionization
fraction can go unphysically negative but this causes no difficulty for
a Boltzmann code and no ambiguity once interpreted as a mode amplitude.

Finally it is interesting to consider the implications for the
total optical depth or equivalently errors on the initial amplitude of
fluctuations $\sigma_{\ln \delta_{\zeta}} \approx \sigma_{\tau}$.  A precise
normalization of the initial conditions is crucial to dark energy
studies utilizing the evolution of structure \cite{Hu02}.
Each principal component mode perturbs 
the total optical depth by 
\begin{equation}
\delta \tau_\mu = \sum_i S_{i\mu} {\delta \tau_i \over \delta x(z_i)}\,.
\end{equation}
This quantity is shown in Fig.~\ref{fig:mode} (lower curve).  The 
best constrained mode bears a $\tau$ uncertainty of $0.0026$.
Note that although the rms uncertainty in the higher modes is large,
the optical depth contributed by them is compensatingly small.  
Given that the total optical depth variation is 
$\delta \tau = \sum_\mu \tau_\mu m_\mu$, the total variance
is given by
$\sigma^2_\tau = \sum_\mu \tau_\mu^2 \sigma_\mu^2.$
We plot in Fig.~\ref{fig:mode} 
the cumulative variance from increasing the number of
modes included in the sum.  The contributions are again dominated 
by the first few modes and placing a prior of $\sigma_\mu < 1$ 
against unphysical
values of $x$ is enough to eliminate the small contribution
from the higher modes.  
Thus, even with an arbitrary ionization history, the total optical depth
can in principle be measured to $\sigma_\tau \approx 0.01$ 
by an ideal experiment.
Note that this is an uncertainty due to cosmic variance and not a bias.
In the example of Fig.~\ref{fig:clmode}, the total optical depth
with 0, 2 and 4 modes is $\tau=0.1732$, $0.1498$, $0.1379$ compared
with a true optical depth of $\tau=0.1375$.
\smallskip

\section{Discussion}

We have studied the effect of an arbitrary ionization history on the
CMB polarization power spectrum and shown that the information
lies in 5 broadly distributed modes in redshift.  We have taken a
complete basis for a range of redshifts $7 < z < 26$ but this can
be extended as needed.  The amplitude of these
modes represent a compact description of the observable properties 
of reionization in the CMB and can be easily predicted  as a filtered version
of any given model for
the ionization fraction.  
They can therefore serve as a model-independent tool 
for data analysis and model testing.  They also can ultimately remove
any bias in the measurement of the total optical depth, or equivalently
the initial amplitude of fluctuations, leaving a residual uncertainty
from cosmic variance at the 1\% level.

\smallskip
{\it Acknowledgments:} We thank A.V. Kravtsov and J. Tumlinson for useful
conversations.
 WH is supported by NASA NAG5-10840 and
the DOE OJI program. GPH is supported by the W.M. Keck Foundation.

\end{document}